\documentclass[aps,prapplied,amsmath,amssymb,amsfonts,reprint,a4paper,superscriptaddress]{revtex4-2}
\usepackage{newtxtext,newtxmath}
\usepackage[T1]{fontenc}
\usepackage[utf8]{inputenx}
\usepackage{graphicx}
\usepackage[dvipsnames]{xcolor}
\usepackage{soul} 
\usepackage[textwidth=17.5cm,textheight=23.5cm,verbose,pdftex]{geometry}
\usepackage[pdftex]{hyperref}
\usepackage{booktabs} 

\hypersetup{pdfauthor={Thomas Auzelle}, bookmarksnumbered=true, pdftitle={Density control of GaN nanowires at the wafer scale using self-assembled SiN$_x$ patches on sputtered TiN(111)}, colorlinks, citecolor=blue, linkcolor=blue, urlcolor=blue}

\newcommand{\etal}[1]{\textit{et al.} [\onlinecite{#1}]}
\newcommand{\degC}{\,$^\circ$C}
\begin{document}
\title{Density control of GaN nanowires at the wafer scale using self-assembled SiN$_x$ patches on sputtered TiN(111)}

\author{T. Auzelle} 
\email[electronic mail: ]{auzelle@pdi-berlin.de}
\author{M. Oliva}
\author{P. John}
\author{M. Ramsteiner}
\author{A. Trampert}
\author{L. Geelhaar} 
\author{O. Brandt} 
\affiliation{Paul-Drude-Institut für Festkörperelektronik, Leibniz-Institut im Forschungsverbund Berlin e.\,V., Hausvogteiplatz 5--7, 10117 Berlin, Germany}

\begin{abstract}
	The self-assembly of heteroepitaxial GaN nanowires using either molecular beam epitaxy (MBE) or metal-organic vapor phase epitaxy (MOVPE) mostly results in wafer-scale ensembles with ultrahigh ($>10$\,\textmu{}m$^{-2}$) or ultralow ($<1$\,\textmu{}m$^{-2}$) densities, respectively. A simple means to tune the density of well-developed nanowire ensembles between these two extremes is generally lacking. Here, we examine the self-assembly of SiN$_x$ patches on TiN(111) substrates which are eventually acting as seeds for the growth of GaN nanowires. We first found that if prepared by reactive sputtering, the TiN surface is characterized by \{100\} facets for which the GaN incubation time is extremely long. Fast GaN nucleation is only obtained after deposition of a sub-monolayer of SiN$_x$ atoms prior to the GaN growth. By varying the amount of pre-deposited SiN$_x$, the GaN nanowire density could be tuned by three orders of magnitude with excellent uniformity over the entire wafer, bridging the density regimes conventionally attainable by direct self-assembly with MBE or MOVPE. The analysis of the nanowire morphology agrees with a nucleation of the GaN nanowires on nanometric SiN$_x$ patches. The photoluminescence analysis of single freestanding GaN nanowires reveals a band edge luminescence dominated by excitonic transitions that are broad and blue shifted compared to bulk GaN, an effect that is related to the small nanowire diameter and to the presence of a thick native oxide. The approach developed here can be principally used for tuning the density of most III-V semiconductors nucleus grown on inert surfaces like 2D materials.
\end{abstract}

\maketitle

Controlling the density of epitaxial GaN nanowire ensembles is a basic requirement for device applications and fundamental studies. In this regard, site-controlled growth of GaN nanowires by deterministic patterning of the substrate surface provides a robust means to tune the nanowire density \cite{Hersee_2006,sekiguchi_2008,bertness_2010,schumann_2011,brubaker_2016,kum_2017}. This process is, however, pricey and fails to produce nanowires with diameters much below $30$\,nm \cite{choi_2012,kano_2015}. Very thin nanowires are still very attractive for achieving pseudomorphic growth of highly mismatched heterostrutures \cite{glas_2006,geng_2012,balaghi_2019} and for growth of quantum dot heterostructures for single photon applications \cite{tribu_2008,reimer_2012}. Such thin nanowires can be obtained by self-assembly on silicon substrates using molecular beam epitaxy (MBE) \cite{stoica_2008,consonni_2010,kaganer_2016}. Yet, in this approach, the growth window to obtain spatially uniform ensembles of long nanowires with low density ($<1$\,\textmu{}m$^{-2}$) is impractically small. It stems from the inability to stop nanowire nucleation before complete coverage of the substrate surface by nanowires \cite{carnevale_2011,zettler_2015,kaganer_2016}. As a result, once reaching the steady-state nanowire elongation regime \cite{fernandez-garrido_2015}, GaN nanowire ensembles self-assembled on Si generally feature densities of $10^2$--$10^3$\,\textmu{}m$^{-2}$ and lower ones are only obtained in microscopic parts of the wafer, where gradients in the substrate temperature or in the atomic fluxes are large \cite{mata_2011,pfuller_2010}. A similar behavior is observed for self-assembled growth on diamond and amorphous AlO$_x$ substrates \cite{schuster_2012,sobanska_2014}. To some extent, the deposition of an AlN buffer layer on Si substrate can help reducing the nanowire density, but a Ga-polar GaN layer also develops in-between the nanowires \cite{bertness_2006,largeau_2012,Chen_2012}. A control on the density of the AlN defects that are seeding the nanowires is also difficult to reach \cite{auzelle_2015a,roshko_2020}. More attractive is the growth on Ti films where $10^1$--$10^2$\,\textmu{}m$^{-2}$ are typically obtained with diameters below $30$\,nm \cite{Wolz_2015,vanTreeck_2018}. Once nitridized, the Ti surface provides a long Ga adatom diffusion length that enables a diffusion-induced repulsion between neighboring nanowires, a process that effectively limits the final nanowire density \cite{vanTreeck_2018}. Following this line, a graphene substrate should provide even longer adatom diffusion lengths to further reduce the density of self-assembled nanowires. Yet, first attempts show that nanowires either nucleate uniformly with densities in the order of $10^2$\,NW\,$\mu$m$^{-2}$ or only at imperfections of the graphene layer such as at step edges \cite{kumaresan_2016,fernandez-garrido_2017a,morassi_2020}. 
Unlike for MBE, metal-organic vapor phase epitaxy (MOVPE) growth of GaN nanowires by self-assembly systematically results in low density ($<1$\,$\mu$m$^{-2}$) with diameters largely exceeding $30$\,nm \cite{koester_2010,tessarek_2013,heilmann_2016}. An approach for the heteroepitaxial growth of thin self-assembled GaN nanowires with densities in the range $10^{-1}$--$10^{2}$\,$\mu$m$^{-2}$ is thus lacking.

To go beyond the limitations given by the direct self-assembly of GaN nanowires, we explore here the self-assembly of nanowire seeds deposited \emph{in situ} prior to the GaN growth. We refrain from using metal droplets acting as seeds in the context of catalyzed growth \cite{chen_2001,wu_2017} to mitigate contamination of the nanowires by the catalyst \cite{cheze_2010} and to avoid issues with the vertical yield \cite{wu_2018}. Instead, we make use of the ability of few Si adatoms to reorganize into small SiN$_x$ patches where GaN nucleation can proceed. This approach has similarities with prior works of Damilano \etal{damilano_2016} and de Souza \etal{desouzaschiaber_2017a} who exploit SiN$_x$ patches that are self-assembled by MBE on GaN(0001) surface for nanowire fabrication. In the first work, the SiN$_x$ patches are used to inhibit thermal decomposition of the underlying GaN. Nanowires are eventually formed by selective area sublimation of the GaN substrate. In the second work, the SiN$_x$ patches are found to alloy with GaN, leading to a local reversal of the crystal polarity from Ga to N-polar above a threshold amount of deposited Si. During GaN regrowth, nanowires form on the N-polar domains whereas a rough layer develops on the Ga-polar areas. 
In both cases, the nanowire density could be varied by more than two orders of magnitude simply by changing the amount of pre-deposited SiN$_x$. However, the main limitation is that the GaN nanowires are obtained on GaN substrates, which prevents, for instance, straightforward access to their optical properties using conventional photoluminescence (PL) spectroscopy. GaN substrates are also more expensive than other conventional substrates such as sapphire or silicon. Finally, very low nanowire densities ($\leq1$\,$\mu$m$^{-2}$) could not be achieved at the wafer-scale. 

Here, we establish a similar process as in Ref.\,[\onlinecite{desouzaschiaber_2017a}] but on TiN(111) films instead of GaN substrates. We first found that if prepared by reactive sputtering, the stoichiometric TiN surface has very low reactivity with GaN, which translates into excessively long incubation times. Fast GaN nanowire nucleation is only induced after annealing of the TiN layer above $1000$\degC{} or after deposition of a sub-monolayer of SiN$_x$. By varying the amount of pre-deposited SiN$_x$, the GaN nanowire density could be tuned by three orders of magnitude with excellent uniformity over the wafer-scale. Importantly, no parasitic GaN growth is observed in-between the nanowires, which allows direct PL characterization of the formed nanowires. At 10\,K, the nanowire band-edge is found to be strongly blueshifted and broad compared to freestanding bulk GaN, both effects that are related to the small nanowire diameter and the presence of a thick native oxide.


\section{Experiments and methods}
TiN substrates are prepared by depositing $200$\,nm of TiN on $2$\,inch Al$_2$O$_3$(0001) substrates using a magnetron sputtering system loaded with a Ti target. The sputtering is performed with an Ar flow of $12$\,sccm and a N$_2$ flow of $3$\,sccm resulting in a total pressure of $10^{-3}$\,mbar. The DC plasma power is set to $560$\,W and a bias of $50$\,V is applied to the substrate. The stoichiometric nature of the TiN is controlled by Raman spectroscopy \cite{Spengler_1978} and spectroscopic ellipsometry \cite{Logothetidis_1996,Logothetidis_1999}. The (111) orientation of the TiN layer is confirmed by x-ray diffraction and reflection high energy electron diffraction (RHEED).
The as-sputtered TiN films are next transferred \emph{in vacuo} to the MBE chamber for GaN nanowire growth. The substrate temperature is determined with an optical pyrometer working at $920$\,nm and the temperature-dependent emissivity of TiN reported in Ref.\,\cite{Briggs_2017}. A correction accounting for light absorption in the view port is applied after calibration of the pyrometer with the $7\times7 \leftrightarrow 1\times 1$ reconstruction transition of a clean Si(111) surface occurring at $\approx 860$\degC{} \cite{Suzuki_1993} and with an emissivity of $0.65$ \cite{Sato_1967}. 
Prior to the GaN growth, a sub-monolayer of Si is deposited on the TiN at $840$\degC{} for $7$\,min using a solid-source effusion cell. The Si flux is calibrated in units of s$^{-1}$\,cm$^{-2}$ by performing secondary ion mass spectrometry measurements of Si-doped GaN layers. Once exposed to active N during GaN growth, the Si is turned into SiN$_x$, for which one bilayer refers here to a density of $1.6\times 10^{15}$\,cm$^{-2}$ \cite{duBoulay_2004}. The N plasma is turned on about $10$\,min after Si deposition meaning that the Si atoms can diffuse on the TiN surface before forming SiN$_x$. For GaN growth, the atomic fluxes are calibrated in GaN-equivalent growth rate units of nm/min by cross-sectional scanning electron microscopy (SEM) of thick GaN layers grown under slightly  N- and Ga-rich conditions on GaN/Al$_2$O$_3$(0001) substrates at $680$--$700$\degC{}. To induce nanowire formation, we use a N flux of $14$\,nm/min ($1.0\times 10^{15}$\,s$^{-1}$\,cm$^{-2}$) provided by a plasma cell and, unless mentioned differently, a Ga flux of $5.5$\,nm/min ($4.0\times 10^{14}$\,s$^{-1}$\,cm$^{-2}$) provided by a solid-source effusion cell. The III/V flux ratio equals $0.4$ for which nanowire nucleation is generally promoted \cite{Fernandez-Garrido_2009}. For samples A--D, different amounts of SiN$_x$ are pre-deposited and the GaN growth duration is set to $155$\,min ($150$\,min for sample B), including the incubation time. We note that a reliable \emph{in situ} measurements of the nanowire incubation time is neither possible here with line-of-sight quadrupole mass spectroscopy (QMS) \cite{fernandez-garrido_2015}, RHEED \cite{Hestroffer_2012}, nor laser reflectance \cite{corfdir_2018} as previously established for GaN nanowire growth on Si substrates. These techniques provide a signal that is proportional to the amount of deposited material and are thus inappropriate for monitoring the growth of very low density nanowire samples ($<10$\,$\mu$m$^{-2}$). 
Once grown, samples are structurally and morphologically characterized by atomic force microscopy (AFM), SEM and transmission electron microscopy (TEM). TEM was performed in a JEOL 2100F field emission microscope operated at $200$\,kV using a multi-beam bright-field mode. Continuous-wave photoluminescence (cw-PL) at $10$\,K is performed in a confocal microscope setup with the $325$\,nm line of an He-Cd laser. 

\section{Results}
\subsection{Impact of TiN surface termination}
\begin{figure*}
	\includegraphics[width = .7\linewidth]{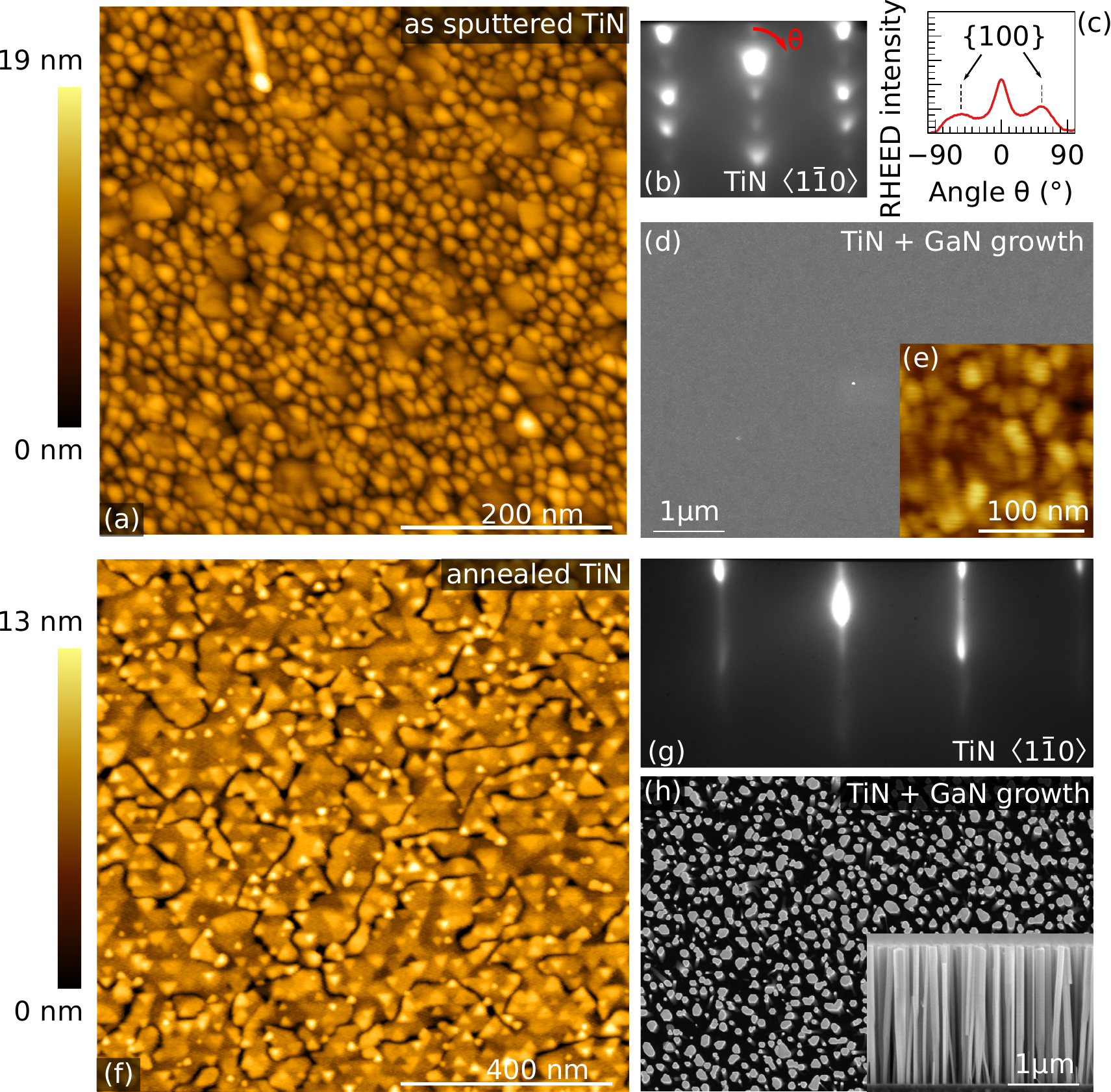}%
	\caption{(a) AFM topograph and (b) RHEED diffraction pattern of an as-sputtered TiN film. The diffraction pattern shows weak chevrons. (c) Circular intensity profile taken around the main diffraction spot in (b), which reveals that the chevrons angle is $55 \pm 5^\circ{}$, in agreement with $\{100\}$ facets. (d) Top view SEM and (e) AFM topographs of the TiN surface after exposure for $5$\,h to active Ga and N beams at $840$\degC{}. The micrographs evidence the absence of GaN nucleation. (f) AFM topograph and (g) RHEED diffraction pattern of a TiN film after annealing in vacuum at $1000$\degC{}. (h) SEM top and side view of the annealed TiN substrate after exposure to active Ga and N beams at $840$\degC{} for $2.5$\,h. The micrographs evidence the nucleation of a high density of GaN nanowires. 
		\label{fig:TiN}}
\end{figure*}

Figure\,\ref{fig:TiN}(a) displays a typical AFM topograph of the as-sputtered TiN layer. The film surface is not flat but composed of small pyramids which results in a spotty RHEED pattern as exemplified in Fig.\,\ref{fig:TiN}(b). Faint chevrons stemming from electron diffraction at the facets of these pyramids can be seen around the RHEED diffraction spots. According to the circular intensity profile taken around the main diffraction spot and shown in Fig.\,\ref{fig:TiN}(c), the angle between the chevrons and the vertical direction amounts to $55\pm5^\circ{}$. This angle is in agreement with the presence of \{100\} facets since they are theoretically tilted by $55^\circ{}$ relative to the (111) plane. Our sputtered TiN films are thus characterized by \{100\} facets, as expected from ab-initio calcultations which predicts a lower energy for the (100) surface compared to the (111) and (110) ones \cite{Marlo_2000}. 

GaN growth is first attempted on the \{100\} faceted TiN film. Remarkably, after $300$\,min of GaN growth using a III/V flux ratio of $0.7$ and a substrate temperature of $840$\degC{}, no GaN nucleation is achieved. This is confirmed \emph{in situ} by the absence of GaN diffraction pattern as seen by RHEED and \emph{ex situ} by the SEM micrograph displayed in Fig.\,\ref{fig:TiN}(d). The substrate surface morphology that is probed by AFM and shown in Fig.\,\ref{fig:TiN}(e) (large scale topograph can be seen in the supplementary information) appears essentially unchanged compared to prior the GaN growth attempt. Yet, a substantial incorporation of Ga in the TiN is deduced from \emph{in situ} measurements of the desorbing Ga flux using QMS (see supplementary information), which is not sufficient to induce GaN nucleation. 
The barrier to the GaN nucleation on the \{100\} faceted TiN surface is only overcome by increasing the III/V flux ratio above $1.3$ 
or by decreasing the substrate temperature below $800$\degC{}, 
both cases eventually resulting in the formation of large, irregular and coalesced GaN crystallites. 

If existing, the growth window to obtain self-assembled GaN nanowires on as-sputtered TiN films appears much smaller than on Ti layers. This is unexpected since for Ti films, the GaN nanowires also nucleate on a thin TiN layer that quickly forms under N exposure of the Ti surface \cite{Wolz_2015,calabrese_2019}. Such nitridized Ti films are, however, characterized by \{111\} facets \cite{calabrese_2019}. In a previous work, we obtained fast nucleation of GaN nanowires on as-sputtered TiN$_x$(111) films by reducing $x$ from $1$ to $0.88$ \cite{auzelle_2021}. According to the RHEED analysis prior to GaN growth, \{111\} facets were also present for these Ti-rich TiN film. It follows that the (111) TiN surface favors GaN nucleation compared to the (100) surface.

To confirm the link between GaN incubation time and TiN surface orientation, we attempt in the following to form \{111\} facets on stoichiometric TiN. This is achieved by annealing the TiN film in the MBE chamber above $1000$\degC{}. At this elevated temperature, the TiN film is seen to undergo a substantial reshaping which results in the formation of flat triangular islands terminated by the (111) plane. This is confirmed by the AFM topograph shown in Fig.\,\ref{fig:TiN}(f) and the related RHEED diffraction pattern given in Fig.\,\ref{fig:TiN}(g). The different in plane orientation of the triangular islands indicates the presence of twin domains which are separated from each other by ridges that are about $20$\,nm wide and $10$\,nm deep. The formation of \{111\} facets is inhibited if the surface is exposed to a flux of active N, which indicates that the reshaping occurs through diffusion of Ti atoms. The onset of thermal decomposition of TiN$_{x<1}$ sub-nitrides \cite{LengauerWalter_1987} or a modification of the surface termination occurring above $1000$\degC{} is proposed to trigger the reshaping. N atoms released from the TiN layer can recombine into volatile N$_2$ molecules, whereas Ti atoms stick to the surface. \{111\} facets would thus be formed as a result of the change in the surface TiN stoichiometry, being energetically favored in Ti-rich conditions. This is in agreement with the fact that \{111\} TiN facets are formed during N exposure of a Ti film \cite{calabrese_2019} and after sputtering of TiN$_{0.88}$ \cite{auzelle_2021}, both cases corresponding to Ti-rich conditions. The TiN reshaping does not induce a substantial change in its Raman signature (see supplementary information), indicating that the change in stoichiometry is restricted to the surface of TiN grains.

GaN growth is further attempted on the annealed TiN film characterized by \{111\} facets, using similar conditions as for the growth on as-sputtered TiN. GaN nucleation is observed already after $10$\,min by RHEED and a well-developed GaN nanowire ensemble is eventually obtained after $2.5$\,h of growth, as exemplified by the SEM micrographs of Fig.\,\ref{fig:TiN}(h). It thus appears that the GaN incubation time on the \{111\} TiN facets is much shorter than on the \{100\} facets, which can be a consequence of the different facet orientation and/or of the different stoichiometry at the TiN surface.
The low reactivity of the \{100\} TiN facets versus GaN nucleation makes the case similar to van-der-Waals epitaxy of GaN nanowires on 2D materials like graphene \cite{chung_2014,heilmann_2016,kumaresan_2016,fernandez-garrido_2017a,liudimulyo_2019,morassi_2020,gruart_2020,barbier_2020}. The inertness of the \{100\} TiN facets is partly preserved after air exposure, since less than $5$\,$\mu$m$^{-2}$ GaN nanowires are obtained on a TiN film stored 40 days in air. 

\subsection{Nanowire seeding using SiN$_x$ patches}
\begin{figure*}
	\includegraphics[width = \linewidth]{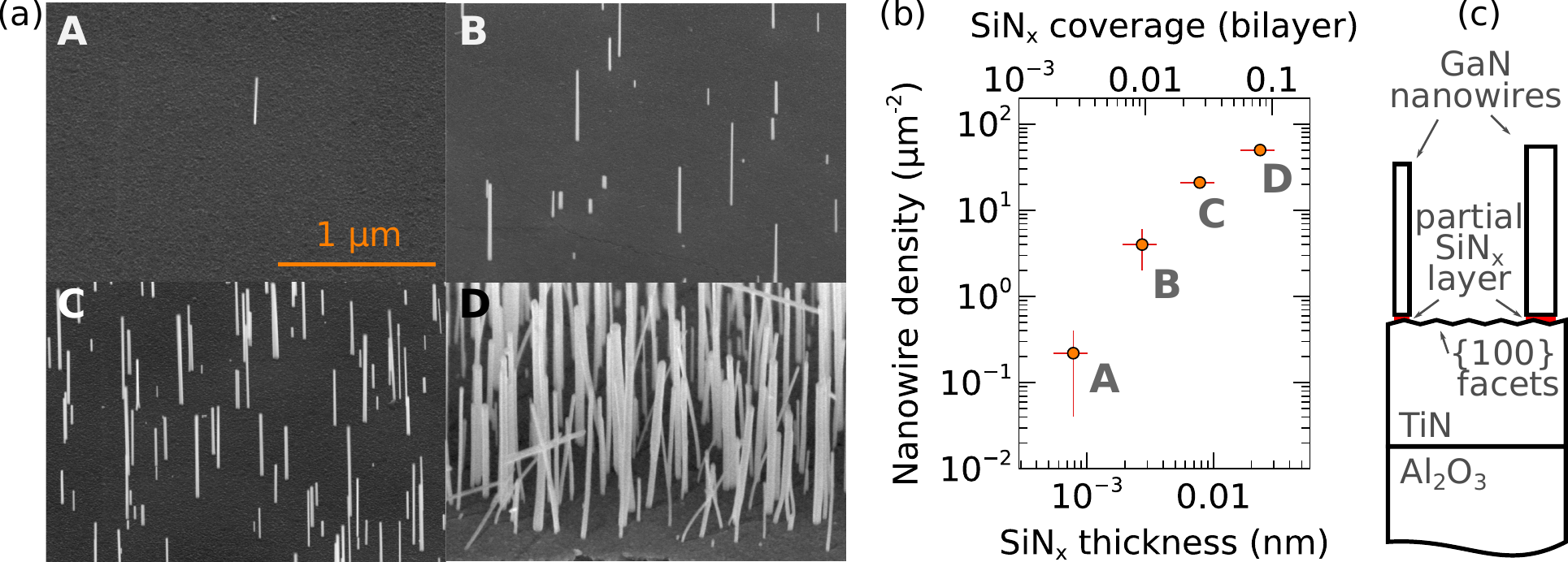}%
	\caption{(a) Bird's eye view SEM micrographs of GaN nanowires grown at $840$\degC{} for 155\,min ($150$\,min for sample B) on as-sputtered TiN with an increasing amount of pre-deposited SiN$_x$. (b) Nanowire density for the samples shown in (a) as a function of the amount of pre-deposited SiN$_x$. The vertical error bar depicts the dispersion in the nanowire density over the $2$\,inch substrate. (c) Schematic of the sample structure. 
\label{fig:growth}}
\end{figure*}

The exceptional inertness of the (100) TiN surface toward GaN nucleation opens perspectives for engineering nanowire nucleation sites. In the following, we seek to create nucleation points on the \{100\} faceted TiN film by depositing a sub-monolayer of SiN$_x$ prior to GaN growth. 
Fig.\,\ref{fig:growth}(a) shows representative SEM images of four nanowire ensembles grown at $840$\degC{} with an increasing amount of pre-deposited SiN$_x$. In each case, nanowires have formed after less than $2$ hours of growth according to RHEED observations. The nanowire density is seen to increase for larger amounts of pre-deposited SiN$_x$, which provides evidence that the presence of SiN$_x$ triggers nanowire nucleation. Fig.\,\ref{fig:growth}(b) indicates the nanowire density as a function of the pre-deposited amount of SiN$_x$ for samples A--D. The vertical error bar relates to the dispersion over the $2$\,inch substrate when neglecting the region 5\,mm away from the substrate rim. It shows that by varying the amount of pre-deposited SiN$_x$, the nanowire density can be changed by almost 3 orders of magnitude with excellent uniformity at the wafer scale. Specifically, this method allows to reach ultralow densities ($<1$\,$\mu$m$^{-2}$), which were previously out-of-reach for GaN nanowires grown by self-assembly by MBE. The proposed mechanism is that Si deposited onto the TiN surface self-assembled into nanometer large round patches, similar to the case of Si deposited on the GaN(0001) surface \cite{desouzaschiaber_2017a}. During GaN growth, the Si patches quickly turn into SiN$_x$ where GaN nanowire nucleation eventually occurs, as schematized in Figure\,\ref{fig:growth}(c).

\begin{figure*}
	\includegraphics[width = .7\linewidth]{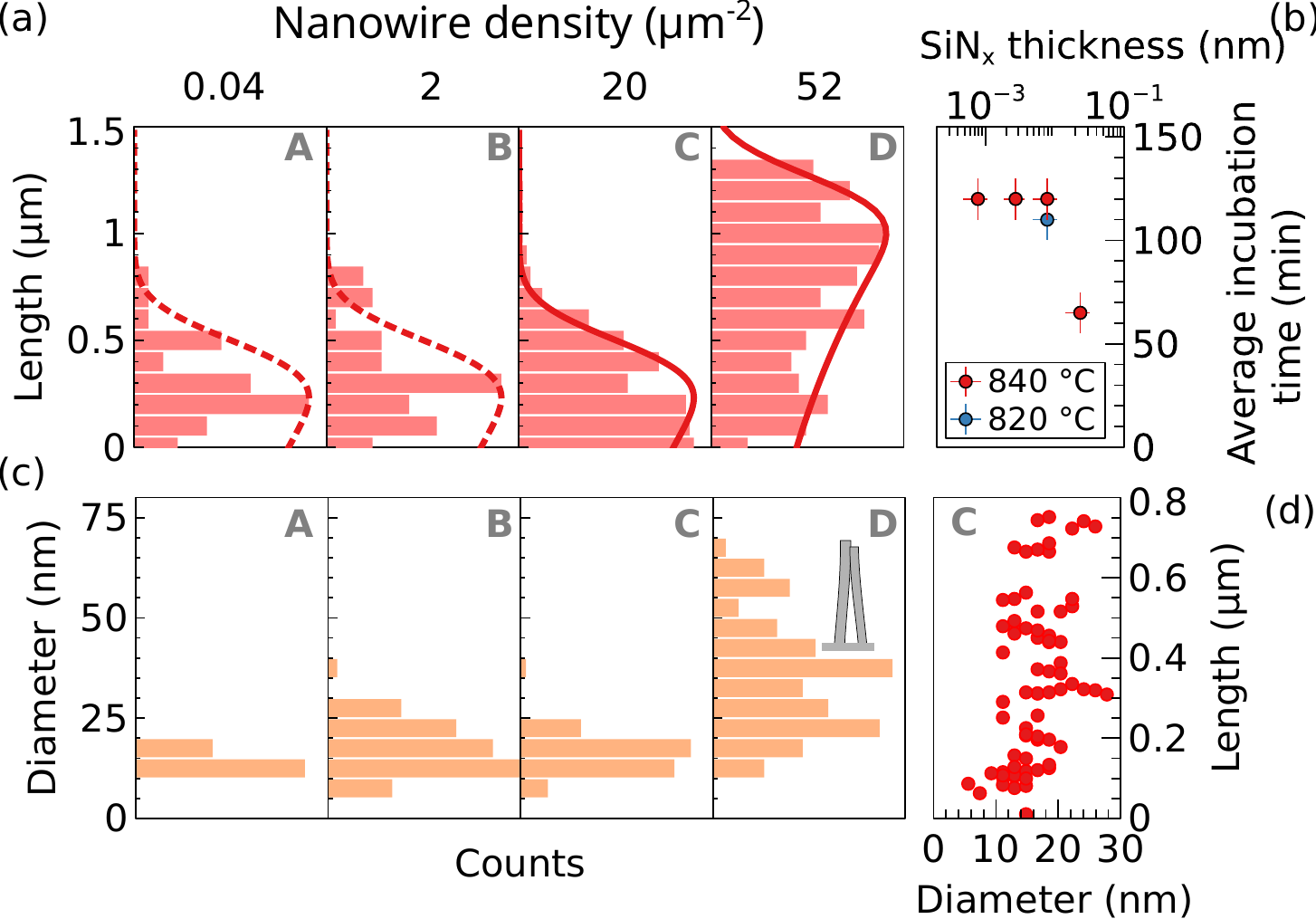}%
	\caption{(a) Length and (c) diameter distributions measured at the wafer center for GaN nanowires grown on TiN with different amount of pre-deposited SiN$_x$. The length histograms are fitted using the nanowire growth model derived in Ref.\,\cite{morassi_2020}. (b) Incubation time for GaN nanowires grown at different temperatures with different amount of pre-deposited SiN$_x$. (d) Length versus diameter measured for single nanowires in sample C. 
		\label{fig:morphology}}
\end{figure*}

To get a better insight into the nucleation mechanism, we scrutinize the morphology of the nanowire ensembles. Fig.\,\ref{fig:morphology}(a) displays the nanowire length distributions for the ensembles A--D at the substrate center. The length distributions are broad, ranging from nearly $0$ to a maximum value of $1.5$\,$\mu$m. We assume that the elongation rate of the nanowire is fixed and equals the N flux ($14$\,nm/min), as expected for thin GaN nanowires grown with a large III/V ratio \cite{consonni_2013,fernandez-garrido_2013}. Under these conditions, a nanowire that would nucleate directly after opening of the Ga and N cell shutters should eventually reach the maximum length of $2.2$\,$\mu$m at the end of the GaN growth. The fact that the observed length distributions are broad with maximum values much smaller than $2$\,$\mu$m suggests the existence of a substantial incubation time that differs for each nanowire \cite{sabelfeld_2013,fernandez-garrido_2015}. Other causes of the broad length distribution could be shadowing between neighboring nanowires and broad diameter distributions \cite{sabelfeld_2013,kaganer_2016}, both of which are neglected here due to the low nanowire density and the consequent absence of pronounced nanowire coalescence. 

As shown in Ref.\,\onlinecite{morassi_2020}, the \emph{average} incubation time $\tau_i$ for a self-assembled GaN nanowire ensemble can be obtained by fitting its length distribution. This approach allows direct comparison of the extracted incubation time for nanowire ensembles of drastically different densities. The length distribution is fitted with the following expression:
\begin{equation}
		f(l,t) = \frac{\alpha}{2A}\exp{\left [ \alpha L+ \frac{\alpha^2\sigma_0^2}{2}\right ]} \text{erfc}{\left [ \frac{L + \alpha \sigma_0^2}{\sigma_0\sqrt(2)}\right ]}
\end{equation}
where $A$ is a normalization factor, $\alpha$ is a dimensionless parameter introduced in Ref.\,\onlinecite{dubrovskii_2016} which describes the ratio between the nucleation probability of the first and subsequent monolayers, $\sigma_0$ is the width of the Gaussian distribution that would be observed for $\alpha = 1$, and $L = l - \nu(t - \tau_i)$ a length expressed in monolayers with $l$ the nanowire length, $\nu$ the steady-state elongation rate of the nanowires (taken here as equal to the N flux), $t$ the total growth duration and $\tau_i$ the average incubation time. The resulting fits for the length distributions of samples A--D are displayed in Fig.\,\ref{fig:morphology}(a). A satisfactory match with the length distributions of ensembles C and D is obtained by taking $\alpha =2\times 10^{-4}$, $\sigma_0 = 580$\,ML and $\tau_i = 120$ and $65$\,min, respectively. These incubation times are consistent with \emph{in situ} observations of the appearance of the first GaN diffraction spots by RHEED. For the ensembles A and B, the ultralow nanowire density prevents measurements of a pool of nanowires that is statistically relevant for fitting the nanowire length distribution. Nevertheless, an acceptable agreement with the experimental length distribution is found when taking the same parameters as used for describing the ensemble C. 

The values of $\alpha$ and $\sigma_0$ obtained here are of the same order of magnitude as reported for the self-assembly of GaN nanowires on graphene patches \cite{morassi_2020}. The average incubation time $\tau_i$ extracted from the above analysis is plotted in Fig.\,\ref{fig:morphology}(b) for ensembles grown with different amounts of pre-deposited SiN$_x$. For comparison, we have also added data for a sample grown at lower temperature ($820$\degC{}) and in otherwise similar conditions. The obtained values in the range of $65$--$120$\,min agree with the extrapolated incubation times given in Ref.\,\onlinecite{fernandez-garrido_2015} for GaN nanowires grown on Si(111) substrates in similar conditions. The long incubation times are then consistent with nanowire nucleation occurring on the SiN$_x$ patches. The observed decrease of the incubation time at lower substrate temperature is also in general agreement with previous works on self-assembled GaN nanowire growth \cite{consonni_2011,fernandez-garrido_2015,sobanska_2016,sobanska_2019,barbier_2020} and results from the reduction in the Ga desorption rate. 
The larger incubation time observed for samples A--C compared to sample D may relate to a smaller thickness of the SiN$_x$ patches in A--C. 

Figure\,\ref{fig:morphology}(c) displays the nanowire diameter distributions extracted from the analysis of bird's eye view SEM micrographs for the ensembles A--D at the substrate center. A narrow distribution at $18 \pm 4$\,nm is observed for samples A--C. In the absence of coalescence, equally thin nanowires are observed for nanowire growth on Si(111) \cite{calarco_2007,kaganer_2016}, on SiN$_x$ patches deposited on GaN(0001) \cite{desouzaschiaber_2017a}, and on Ti(0001)\,\cite{vanTreeck_2018}. The diameter distribution only broadens toward larger diameters for sample D. According to Ref.\,\onlinecite{kaganer_2016}, the critical length above which bundling between two nanowires of diameter $18$\,nm becomes energetically favorable is $0.5$ and $0.4$\,$\mu$m for the nanowire density of ensembles C and D, respectively. Hence, unlike in sample C, most nanowires from sample D are expected to have coalesced during growth due to bundling, which can explain the presence of thick nanowires. 

Figure\,\ref{fig:morphology}(d) shows the diameter versus length for nanowires from sample C. No correlation between these two quantities can be seen, meaning that nanowires do not get thicker during their elongation stage. %
The diameter distribution in the ensembles A--C thus likely reflects the size of the SiN$_x$ patches on top of which the nanowires have nucleated. 

\begin{figure}
	\includegraphics[width = \linewidth]{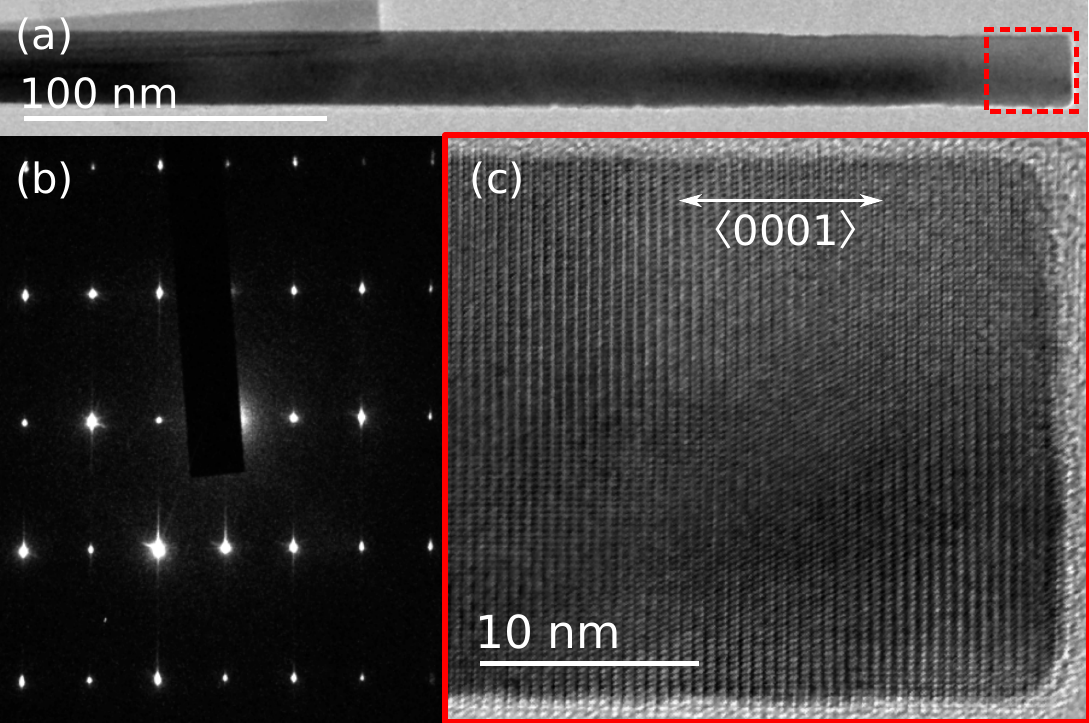}%
	\caption{(a) Bright-field transmission electron micrograph of a single representative GaN nanowire from sample C dispersed on a carbon lacey grid. (b) Selective area electron diffraction pattern of the nanowire shown in (a). (c) Lattice image of the tip of the nanowire shown in (a) taken along the $\langle 11\overline{2}0\rangle$ zone axis.
		\label{fig:TEM}}
\end{figure}

TEM of ten nanowires was performed to assess the crystalline structure of the nanowires. A bright-field image of a representative single nanowire from sample C is shown in Fig.\,\ref{fig:TEM}(a). It reveals the absence of extended defects such as dislocations, inversion domains or the presence of a high density of basal-plane stacking faults. The selective area electron diffraction pattern and the high-resolution micrographs of Fig.\,\ref{fig:TEM}(b) and \ref{fig:TEM}(c) confirm the wurzite structure of the nanowire, with the $\langle0001\rangle$ direction parallel to the nanowire axis. Such a nanowire morphology is very similar to that of uncoalesced N-polar GaN nanowires grown on other substrates such as Si or TiN\{111\}.

In conclusion, the presence of the SiN$_x$ patches is seen to enable GaN nanowire nucleation on TiN\{100\} by providing a nucleation site with properties similar to the surface of bulk Si(111) substrates. We thus expect GaN nanowire nucleation on the SiN$_x$ patches to occur in a comparable fashion as for growth on Si(111) \cite{consonni_2013}. The very different temporal evolution of nanowire nucleation on the TiN\{100\} and SiN$_x$ regions is ultimately related to their different intrinsic incubation times. A phenomenological description of the latter could be found in the theoretical framework of burst nucleation \cite{mer_1952}. In such a model, the longer incubation time on TiN\{100\} compared to SiN$_x$ could be associated to a shorter residence time of the Ga adatoms on TiN\{100\}, or to the need for a larger critical size of GaN nuclei due to the high energy of the incoherent GaN/TiN\{100\} interface.

\subsection{Nanowire luminescence}
\begin{figure*}
	\includegraphics[width = .9\linewidth]{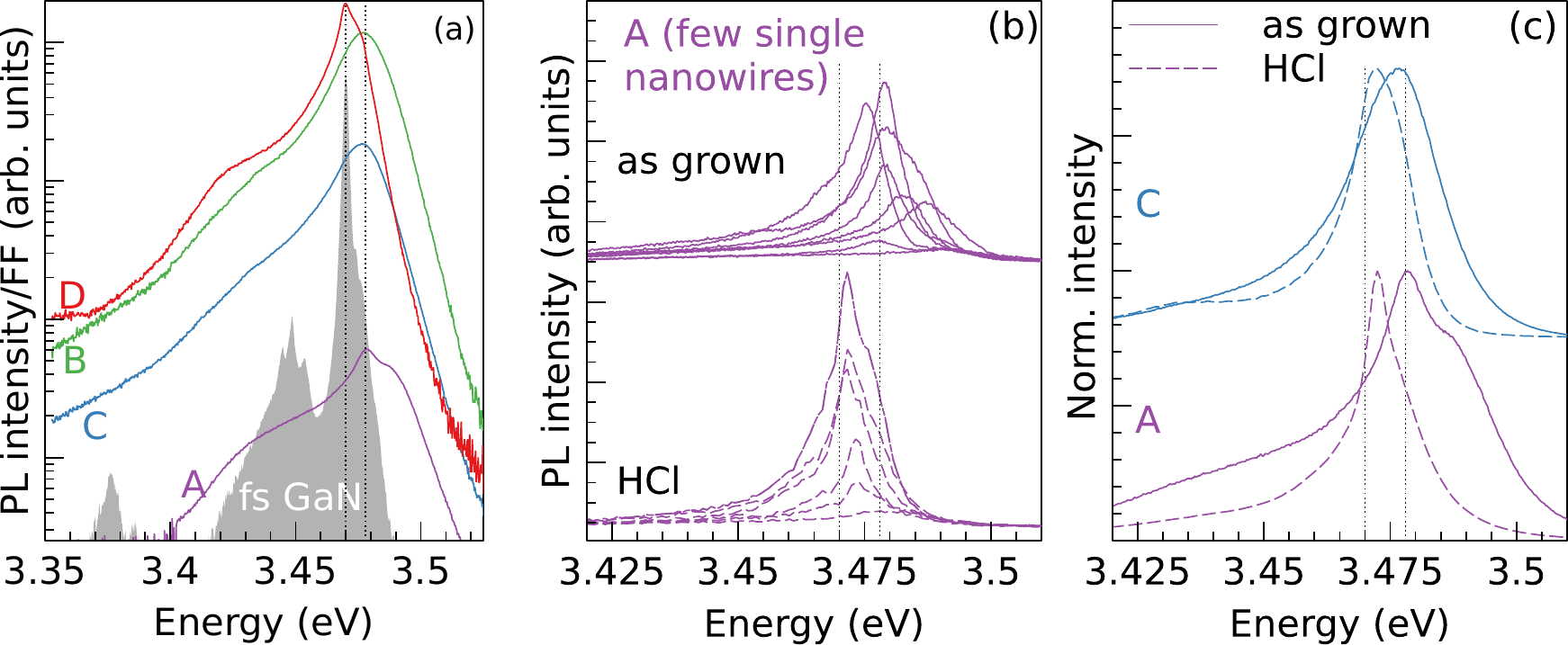}%
	\caption{(a) cw-PL spectra of GaN nanowires grown on TiN and of freestanding GaN. The PL intensity is divided by the nanowire fill factor (FF) to ease comparison between nanowire ensembles of drastically different densities. (b) Several cw-PL spectra of a few single nanowires taken from sample A and measured before and after HCl etching. Spectra are vertically shifted for clarity. (c) Normalized cw-PL spectra of the ensembles A and C before and after HCl etching. Spectra are vertically shifted for clarity. All PL acquisitions in (a)--(c) are done at 10\,K, with a laser spot size of $4$\,$\mu$m$^2$, and an excitation density of about $2$\,kW/cm$^2$. In (a) and (c), spectra from sample A have been spatially averaged over a sample area $>10^3$\,$\mu$m$^2$. The vertical dotted lines indicate the expected transition energies for the D$^0$X and the X$_A$ in bulk GaN at $3.472$ and $3.478$\,eV, respectively \cite{Monemar_2010}.
		\label{fig:PL}}
\end{figure*}

Figure\,\ref{fig:PL}(a) compares the cw-PL spectra acquired at 10\,K of the nanowire samples A--D and of freestanding GaN (fs GaN). To ease the comparison between nanowire ensembles of drastically different densities, their PL intensity is divided by the nanowire surface fill factor (FF). In spite of this correction, sample A --- with the lowest nanowire density --- has a much lower PL intensity than other samples, which is associated to an inefficient coupling of the $325$\,nm laser light into the nanowires. Indeed, by modeling the laser as a plane wave incoming parallel to the substrate surface, light absorption in vertical GaN nanowires with diameter below $50$\,nm becomes negligibly small \cite{Hauswald_2017}. Larger absorption can however be obtained by increasing the component of the laser polarization axis that is parallel to the nanowire axis. Following this argument, we thus explain the higher PL intensities of samples B--D to the fact that the laser light can first be scattered by nanowire tips before being absorbed at the sidewalls of neighboring nanowires. 

The nanowires feature an excitonic band edge luminescence that is broader ($25$\,meV for sample A) and blueshifted ($3$\,meV for sample A) compared to the fs GaN, an effect that relates to their very small diameters. In short, donors located close to the surface exhibit a different binding energy and a different wavefunction symmetry compared to in the bulk \cite{corfdir_2012}. The dielectric mismatch between air and GaN also affects the exciton binding energy and oscillator strength in thin nanowires \cite{corfdir_2015,zettler_2016}. Additionally, for diameters below $10$\,nm, radial confinement sets in \cite{brockway_2011}. Last, the surface stress results in a significant homogeneous strain already for nanowire diameters below $50$\,nm\,\cite{Calabrese_2020}. As a result of all of these contributions, both the free and donor-bound exciton in GaN nanowires show a significant blueshift in the PL signal when reducing the nanowire diameter below $\approx 30$\,nm \cite{corfdir_2016,zettler_2016}. This energy shift scales nonlinearly with the diameter, which means that a significant broadening of the band edge luminescence is expected even if measuring a nanowire ensemble with a narrow diameter distribution as having here for sample A--D. 

The average distance between neighboring nanowires of sample A is on the same order of magnitude as the laser spot size ($\approx 4$\,$\mu$m), which allows measuring a few single nanowires only. Exemplary PL spectra of these nanowires are shown in Fig.\,\ref{fig:PL}(b). The spectra feature $\approx10$\,meV broad PL lines centered between $3.475$ and $3.490$\,eV. The variability in the diameter of the measured nanowires can account for the spread in PL energies observed here, but not for the substantial linewidths. In comparison, single freestanding nanowires with diameters above $20$\,nm generally exhibit sharper lines with linewidth below $2$\,meV \cite{pfuller_2010,Chen_2011,gorgis_2012,Avit_2014,Mancini_2019}.

Broad band edge PL transitions are typically associated with a large amount of inhomogeneous strain and/or to a large density of ionized impurities. X-ray measurements of the nanowire ensemble C, however, reveal a homogeneous strain of $10^{-4}$ and a microstrain of $3\times 10^{-4}$, which should lead to a broadening not exceeding $5$\,meV \cite{Fernandez-Garrido_2014}. In a recent work, we have also estimated a donor concentration $<10^{17}$\,cm$^{-3}$ for GaN nanowires grown on TiN$_x$ \cite{auzelle_2021}, which is comparable to GaN nanowires grown on Si \cite{calarco_2005} and for which sharp transitions are observed \cite{pfuller_2010}. Hence, we propose that the unusually large PL linewidth observed here for single GaN nanowires relates to the presence of a thick and disordered native oxide, which builds inhomogeneous strain and/or induces a large potential fluctuation due to a random distribution of ionized states. The presence of a few nanometer thick oxide shell has already been proposed to account for the roundish shape of many GaN nanowires grown on Ti films, which contrasts with the hexagonal shape of most GaN nanowires grown on Si substrates \cite{Kaganer_2022}. The oxide is believed to form as a result of massive Ga incorporation in the Ti layer during exposure to the Ga flux \cite{calabrese_2019}. Once the sample is cooled down, part of the incorporated Ga is released, wets the GaN nanowire sidewalls and is turned into GaO$_x$ during air exposure. 

Here, a massive incorporation of Ga in the sputtered TiN layer is also observed to occur during exposure to the Ga flux (see supplementary information). To remove the GaO$_x$ shell that has potentially formed, ensembles A and C are dipped at room temperature for $2$\,min in a $30\%$ aqueous HCl solution and their PL spectra before and after etching are compared in Fig.\,\ref{fig:PL}(b) and \ref{fig:PL}(c). Remarkably, the surface treatment has induced a substantial redshift and narrowing of the PL lines, which is also visible in the spectra of a few single nanowires plotted in Fig.\,\ref{fig:PL}(b). These results are consistent with a partial removal of a thick oxide shell which would introduce a tensile strain along the nanowire length. The HCl treatment is also seen to break most of the nanowires. As discussed in Ref.\onlinecite{oliva_2022}, the capillary forces acting at the nanowire tips when drying the sample in air are strong enough to induce buckling and breaking. Most of the measured nanowires after HCl etching are thus lying on the substrate, which enhances the coupling with laser light but may also add an inhomogeneous strain when cooling the substrate down to cryogenic temperature for PL measurements \cite{Anufriev_2012}. The latter can account for the PL linewidths of the single nanowires that are still larger than $2$\,meV in spite of the GaO$_x$ removal. The observation of sharp transitions in ultrathin GaN nanowires on TiN will then require optimizing experimental methods to get rid of the native oxide and to enhance light coupling.


\section{Summary and conclusions}
The incubation time of GaN nanowires grown on sputtered TiN(111) was seen to exceed 5 hours for the \{100\} terminated surface, which documents the extraordinary inertness of these facets with regard to GaN nucleation. GaN nucleation is eventually obtained by self-assembling SiN$_x$ patches on the TiN surface prior to GaN growth. This process provides a precise control on the nanowire density at the wafer scale, simply by tuning the amount of pre-deposited SiN$_x$. Specifically, ultralow nanowire densities as typically obtained by direct nanowire self-assembly with MOVPE are now in reach with MBE. The characterization of the nanowire ensemble morphology agrees with GaN nucleation occurring on SiN$_x$ patches with a comparable incubation time as for GaN nanowire growth on Si(111) substrates. The broad length distribution for the samples examined here mainly results from the substantial incubation time ($\tau_i>60$\,min). This effect could be principally mitigated by performing a two-step growth \cite{carnevale_2011,zettler_2015} or simply by using seeds of different nature, like AlN as reported for GaN nanowire growth on graphene \cite{liudimulyo_2020}. The PL spectra of single freestanding nanowires are dominated by donor-bound and free excitons. These transitions are broader and occur at higher energies compared to the bulk, an effect that is related to the small nanowire diameter and to the presence of a thick native oxide. 

The new ability to self assemble at the wafer scale ultrathin GaN nanowires with densities below $1$\,$\mu$m$^{-2}$ by MBE is unprecedented and opens new growth opportunities. An attractive example is the realization of coherent highly-mismatched core-shell nanowire heterostructures \cite{balaghi_2019} that would not suffer from inhomogeneity due to shadowing between neighboring nanowires \cite{hestroffer_2013,Schroth_2019}. More generally, the approach developed here for tuning the density of GaN nuclei on the inert TiN surface may prove useful in the context of van-der-Waals epitaxy of III-V on 2D materials. For instance, a precise control of the density of nucleation sites is critical to obtain smooth and epitaxial layers \cite{Wu_2017a} while allowing exfoliation of the layers at a later stage \cite{vuong_2020}.

\section{Acknowledgment}

We are indebted to Carsten Stemmler and Katrin Morgenroth for their dedicated maintenance of the molecular beam epitaxy system. We thank Neha Aggarwal for a critical reading of the manuscript. Funding from the Deutsche Forschungsgesellschaft through project AU 610/1-1 is gratefully acknowledged.

\bibliography{GaNNWs_TiN_density}

\end{document}


\title{Density control of GaN nanowires at the wafer scale using self-assembled SiN$_x$ patches on sputtered TiN(111) \\ Supplementary information}

\author{T. Auzelle} 
\email[electronic mail: ]{auzelle@pdi-berlin.de}
\affiliation{Paul-Drude-Institut für Festkörperelektronik, Leibniz-Institut im Forschungsverbund Berlin e.\,V., Hausvogteiplatz 5--7, 10117 Berlin, Germany}

\author{M. Oliva}
\author{P. John}
\author{M. Ramsteiner}
\author{A. Trampert}
\author{L. Geelhaar} 
\author{O. Brandt} 
\affiliation{Paul-Drude-Institut für Festkörperelektronik, Leibniz-Institut im Forschungsverbund Berlin e.\,V., Hausvogteiplatz 5--7, 10117 Berlin, Germany}


\maketitle

\subsection{TiN annealing}

\begin{figure}[h]
	\includegraphics[width = \linewidth]{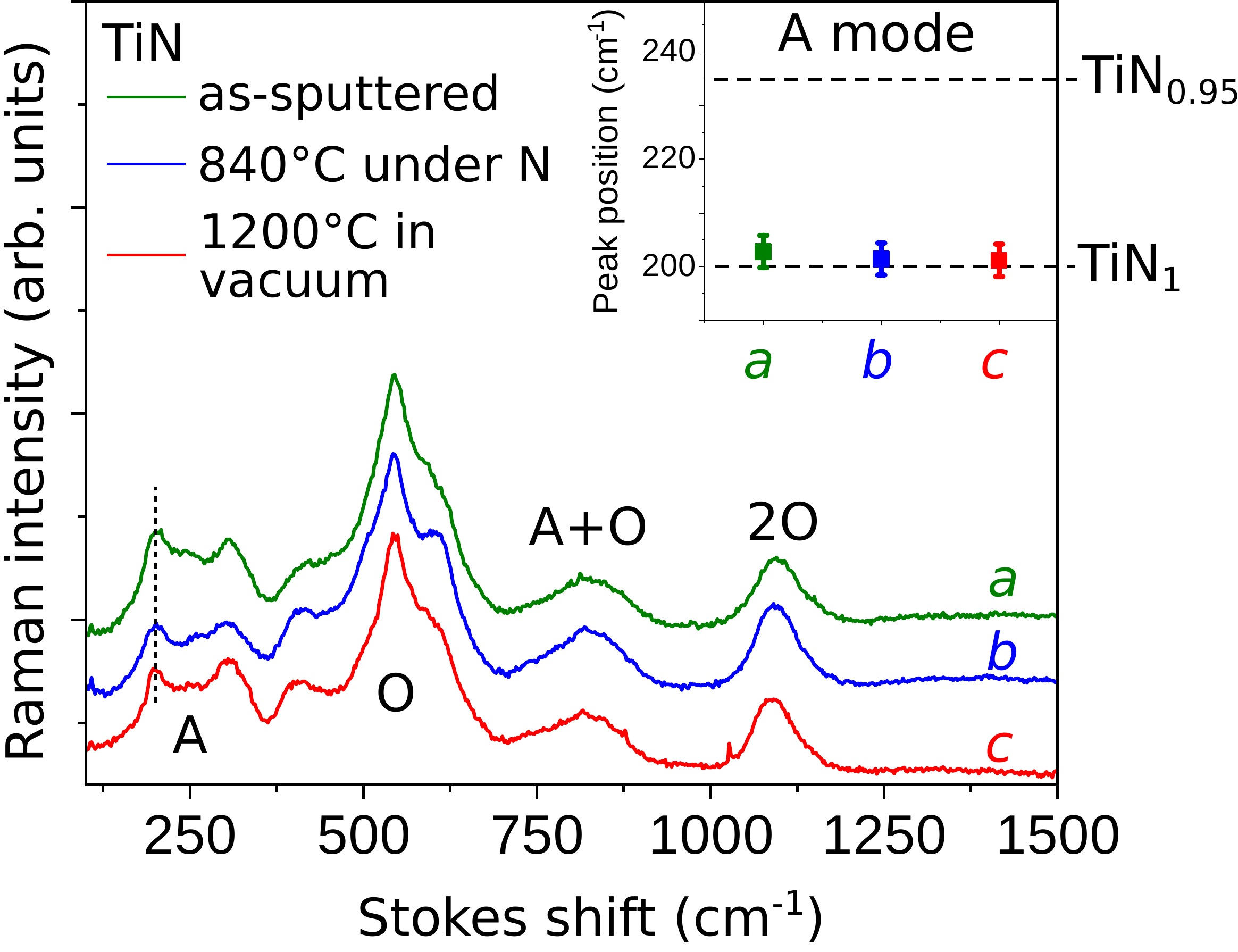}%
	\caption{Raman spectra of a non-annealed TiN film, after TiN annealing at $840$\degC{} under active N exposure, and after TiN annealing at $1200$\degC{} in vacuum. Spectra are vertically shifted for clarity. The various TiN acoustic (A) and optic (O) modes of first and second orders are shown according to Ref.\,\onlinecite{Spengler_1978}. The inset shows the wavenumber of the first acoustic mode for the three films measured here in comparison to the expected value for stoichiometric TiN and TiN$_{0.95}$. 
		\label{fig:Raman}}
\end{figure}

A possible modification of the stoichiometry of the TiN films during annealing is examined by Raman spectroscopy performed at room temperature with a laser at $473$\,nm. Fig.\,\ref{fig:Raman} shows the Raman spectra of a non-annealed TiN film, after annealing under active N exposure, and after annealing in vacuum. The various acoustic and optical modes appear unaffected by the high temperature annealing. In addition, the first order acoustic mode is at $202\pm2$\,cm$^{-1}$ for the three TiN films, which is consistent with stoichiometric TiN \cite{Spengler_1978}. Thus, the TiN reshaping occurring during vacuum annealing above $1000$\degC{} is not caused by a modification of the bulk TiN stoichiometry. A possible modification of the TiN surface stoichiometry would not be seen here.

\subsection{TiN properties after GaN growth}

\begin{figure*}
	\includegraphics[width = \linewidth]{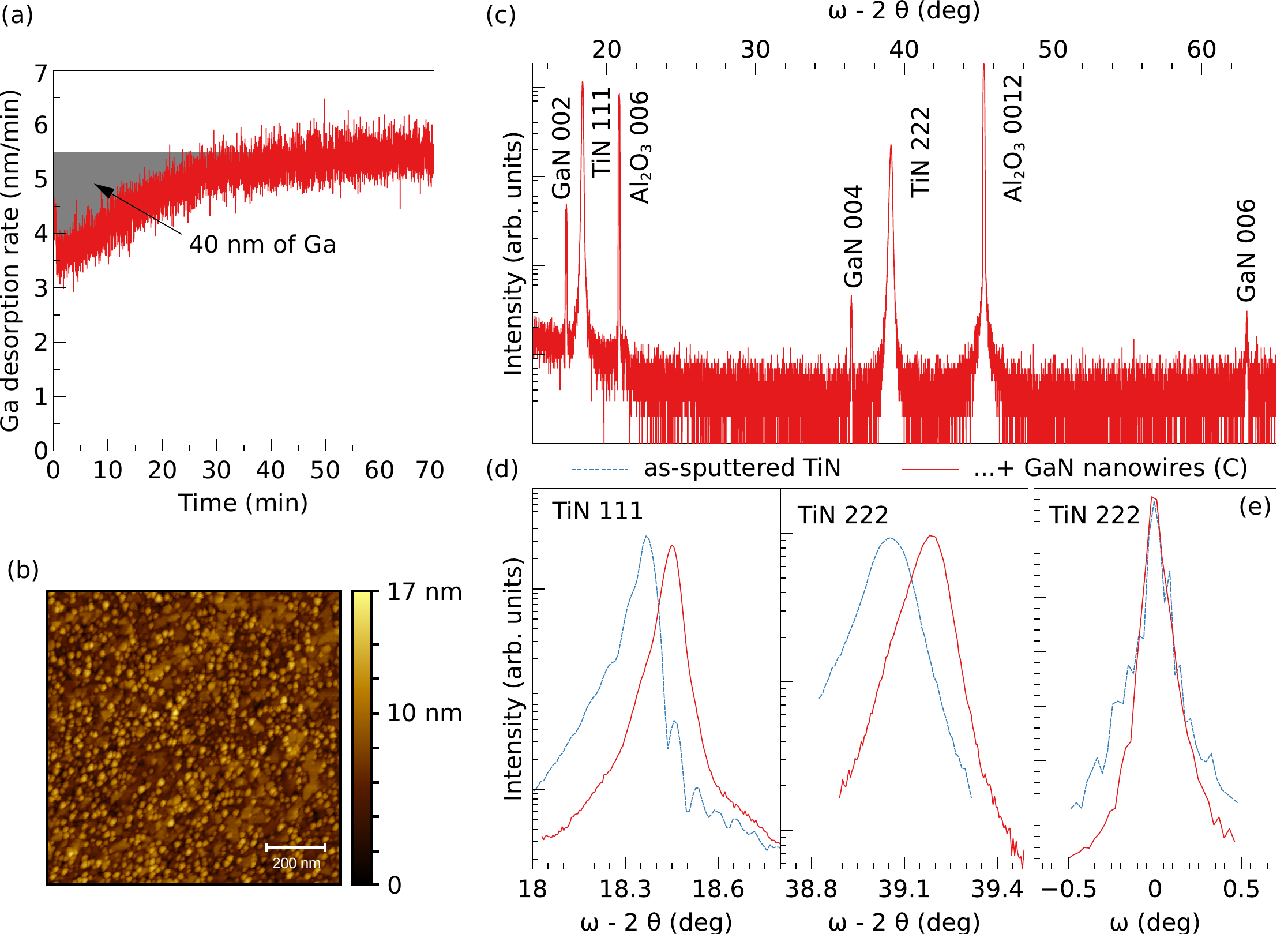}%
	\caption{(a) Flux of desorbing Ga measured by line-of-sight quadrupole mass spectroscopy (LoS-QMS) during the first $70$\,min of GaN nanowire growth on as-sputtered TiN. For that sample, the incubation was exceeding $70$\,min. (b) Atomic force topograph of the TiN surface after $5$h exposure to the N and Ga beams at $840$\degC{}. (c)--(d) Symmetric $\omega$-$2\theta$ and (d) $\omega$ XRD scans of the as-sputtered TiN substrate and after GaN nanowire growth (sample C). 
		\label{fig:TiN_aging}}
\end{figure*}

The properties of the TiN film before and after GaN growth are investigated to check for a possible aging of the film. In particular, line-of-sight quadrupole mass spectroscopy (LoS-QMS) is used to measure the amount of desorbing Ga during the first minutes of GaN growth. As shown in Fig.\,\ref{fig:TiN_aging}(a), a substantial amount of Ga is seen to incorporate in the sample before completion of the GaN incubation time. The final concentration of Ga in the $200$\,nm thick TiN layer is about $20$\%. A very similar phenomenon is observed for GaN growth on a Ti film and is associated to the incorporation of Ga in the Ti layer, forming a GaTi alloy \cite{calabrese_2019}. Here, the solubility of Ga in stoichiometric TiN is considered negligible and we propose that the Ga fills instead defect sites like twin domain boundaries. Several studies have shown that sputtered TiN films are typically porous (``spongelike'') due to a presence of tiny gaps between different TiN grains \cite{Park_1996}. As a result, large quantities of Al or O can accumulate in the TiN film \cite{Mandl_1990,Eizenberg_1994}, up to $20$\% for the films of lowest density. Reduced Ga incorporation could thus be achieved by improving the crystalline quality of the TiN film.

Figure\,\ref{fig:TiN_aging}(b) shows an atomic force topograph of a TiN film after $5$\,hours of exposure to the Ga and N beams at $840$\degC{} and in the absence of Si pre-deposition. No GaN nucleation can be observed and the TiN surface appears essentially unchanged compared to the as-sputtered one in spite of the substantial Ga incorporation discussed previously. 

Last, Fig.\,\ref{fig:TiN_aging}(c) shows the symmetric $\omega$/$2\theta$ X-ray diffraction profile acquired over a large angular range for the GaN nanowire ensemble C. Diffraction lines from the Al$_2$O$_3$, wurtzite GaN and $\delta$-TiN can be seen, corresponding to the substrate, the GaN nanowires, and the TiN film, respectively. No signature from the sub-phase $\epsilon$-Ti$_2$N is observed. The 111 and 222 diffraction lines of TiN measured before and after GaN growth are compared in Fig.\,\ref{fig:TiN_aging}(d). These lines are seen to shift to higher angles after GaN growth, giving a relative change in the d$_{111}$ spacing of about $-0.3$\%. This is associated to strain build by the incorporation of Ga in the TiN layer. We note that the width of the diffraction lines remains essentially unchanged after GaN growth, which confirms the absence of degradation of the bulk of the TiN grains. Similarly, the $\omega$ scans of the TiN diffraction line acquired before and afer GaN growth and shown in Fig.\,\ref{fig:TiN_aging}(e) do not indicate an increased crystalline disorder after Ga incorporation.

\bibliography{GaNNWs_TiN_density}